\documentclass[sigconf]{acmart}
\AtBeginDocument{%
  \providecommand\BibTeX{{%
    \normalfont B\kern-0.5em{\scshape i\kern-0.25em b}\kern-0.8em\TeX}}}

\usepackage{pdfpages}
\usepackage{soul}
\usepackage{appendix}
\usepackage{graphicx}
\usepackage{caption}
\usepackage{subcaption}

\begin{document}

\title{Exploring post-neoliberal futures for managing commercial heating and cooling through speculative praxis}

\author{Oliver Bates}
\email{o.bates@lancaster.ac.uk}
\affiliation{%
  \institution{Lancaster University}
  \country{UK}
}
\email{oliver@fractals.coop}
\affiliation{%
  \institution{fractals co-op}
  \country{UK}
}

\author{Christian Remy}
\email{c.remy@lancaster.ac.uk}
\affiliation{%
  \institution{Lancaster University}
  \country{UK}
}

\author{Kieran Cutting}
\email{kieran@fractals.coop}
\affiliation{%
  \institution{fractals co-op}
  \country{UK}
}

\author{Adam Tyler}
\email{adam.tyler@lancaster.ac.uk}
\affiliation{%
  \institution{Lancaster University}
  \country{UK}
}

\author{Adrian Friday}
\email{a.friday@lancaster.ac.uk}
\affiliation{%
  \institution{Lancaster University}
  \country{UK}
}

\renewcommand{\shortauthors}{Bates, et al.}

\begin{abstract}
What could designing for carbon reduction of heating and cooling in commercial settings look like in the near future? How can we challenge dominant mindsets and paradigms of efficiency and behaviour change? How can we help build worlds through our practice that can become future realities? This paper introduces the fictional consultancy ANCSTRL.LAB to explore opportunities for making space in research projects that can encourage more systems-oriented interventions. We present a design fiction that asks `what if energy management and reduction practice embraced systems thinking?'. Our design fiction explores how future energy consultancies could utilise systems thinking, and (more than) human centred design to re-imagine energy management practice and change systems in ways that are currently unfathomable. We finish by discussing how LIMITS research can utilise design fiction and speculative praxis to help build new material realities where more holistic perspectives, the leveraging of systems change, and the imagining of post-neoliberal futures is the norm. 
\end{abstract}





\setcopyright{none}
\settopmatter{printacmref=false}
\acmDOI{}
\acmISBN{}
\acmConference[LIMITS '24]{Tenth Workshop on Computing within
Limits}{June
18--19, 2024}{Online}

\maketitle

\section{Introduction}

Our current project, Net0i\footnote{\url{http://net0i.org/}}, looks at developing tools to help stakeholders with their energy control and management practices in commercial settings.  These practices consist of analysing energy data to make decisions that reduce energy consumption and the carbon footprint of the organisations they work for. Our mission is research that helps stakeholders achieve real impact and initiate a transition towards a Net Zero future. Our current project focuses on the opportunities for ICT to reduce the energy consumption of heating and cooling buildings \cite{tyler2024}, and making more of energy data through better contextualisation in commercial settings \cite{remy2024wasted}. Whilst working towards this mission we have found that systems change sits in conflict with the current configuration of energy management practices. We believe it is imperative that the full range of stakeholders look at energy and carbon reduction through a systems thinking lens that understands the complexity of a system, rather than a focus on blunt interventions that seek to reduce energy consumption metrics. 

Energy management is enacted through organisational views (e.g., interactive data dashboards, spreadsheets, policies) with limited ability to intervene or directly influence the ``\textit{local populations}'' occupying physical spaces~\cite{goulden2015caught}.
The spreadsheet and dashboard are two of the main sites in which energy management is performed and through which optimisations are sought. These dashboards are designed to help increase the efficiency of the existing energy system, but due to their limited view they act as a barrier to interventions that consider behaviours and interactions of complex systems (people, businesses, environment). Furthermore, those current lenses focus on existing infrastructure and technology, neglecting and ignoring large parts of the organisational system, such as purpose and human experience, comfort, and well-being. Beyond this, commercial energy infrastructure is controlled through interfaces to Building Management Systems (BMS) which allow digital monitoring and control of energy flows. 

Given current tools and established practices it is not surprising that systems thinking can be `out of scope' in energy management practice; a BMS provides the tools to change settings of the buildings, not in relation to the people in it \cite{remy2024wasted}. It provides one view on the organisation as an array of metrics to be managed and optimised. These practices are grounded in an engineering-led practice that relies heavily on analysing data to find `opportunities for optimisation'. This is done through the analysis of quantitative data, such as temperatures, air flow rates, humidity measures, air quality parameters, and other data sources from commonly installed sensors. These practices and tools can make it more difficult to imagine energy policy change~\cite{new2023space}, and hide where organisations' policies or missions have implications for energy use and optimisation~\cite{gormally2019doing}.

When thinking about the possible ways to reduce energy consumption, an organisation's energy analysis will often be reduced down to KPIs and metrics that can be tracked against; completely decontextualising people, the organisational goals and systems implicated in energy demand or production. We feel strongly that tools like dashboards and spreadsheets, as well as the suite of analytical approaches that energy managers often deploy, limit their ability to make changes that have systemic and thus significant outcomes. The solution cannot be an `incremental data-driven fix' within the current tools, as the tools do not allow for considering inputs and outputs of a system beyond energy, such as taking into account the diverse ways people use buildings into account and the underlying systemic drivers for this.

In this paper, we challenge the existing resources and practices of commercial energy management and present a fundamentally different way of thinking about energy management. We present a playful and goal-driven design fiction of what could be, based on the experiences from our work in the domain. Our approach does not directly critique the concrete practices we have been involved with, but instead uses artistic freedom to allow us to be more candid in our assessment of `what currently is', yet at the same time go one step further in `what could be'. We argue why our design fiction approach helps to enable out of the box thinking and what it offers for the energy management design space, and discuss our designs in light of potential future interventions to inspire researchers within the LIMITS community and beyond to implement tools that reshape energy management through different political-economic lenses (e.g., moving from neoliberal individualism and optimisation to something post-neoliberal).



\section{Why Design Fiction?}

Over several years, we have worked with numerous stakeholders in the realm of energy management. Among the most commonly identified patterns we have witnessed is the laser sharp focus on data-driven energy savings; be it in analysing statistical anomalies and patterns, investigating leaks or gaps in data and the infrastructure it was sourced from, or reducing demand by adjusting settings of centralised or distributed systems~\cite{bates2017beyond, gormally2019doing}. The goal is to save energy---with the overarching aim to ultimately save money (and in many cases, with becoming more sustainable only as a fortunate side effect).

Our project aimed to investigate this process and, based on insights gained from data, provide pointers for interventions towards a Net Zero future. The biggest insight that we found, however, was that the current system---the interplay between infrastructure (from organisations and buildings down to the pipes, wiring, and sensors), people, and organisation practice---does not allow for anything but incremental changes (cf. \cite{remy2024wasted}). Changes in complex systems are understood to happen by the strategic choice of a ``\textit{leverage point}'': places where ``\textit{a small shift in one thing can produce big changes in everything}''~\cite[p. 1]{meadows1999leverage}. The leverage points that are suggested by everyday practice and technologies in the energy management system are those that are least effective for engendering change: what Donella Meadows describes as ``\textit{constants, parameters, numbers}'', and ``\textit{the sizes of buffers}''~\cite[p. 2]{meadows1999leverage}. To work towards a Net Zero future, energy management practice needs a complete overhaul---which requires reaching towards leverage points that are more effective at creating systems change. In this specific case, we suggest that identifying ways to change ``\textit{the rules of the system}'', ``\textit{the goals of the system}'', and ``\textit{the mindset or paradigm of out which the system\ldots arises}'' could be most effective for movement towards a net zero or carbon negative future~\cite[p. 3]{meadows1999leverage}.

The fact that radical change is needed is not a new insight---it is a commonly known that the best way to increase the energy efficiency of many buildings is to tear it down and start again from scratch, or at the very least do a likely expensive large-scale renovation bringing the insulation up to most recent standards. But this is the work of engineers, not energy managers, data scientists, or researchers, and thus out of our work's scope as well as that of our project's affiliated stakeholders. If we know that such a fundamental change is needed, but the most obvious changes are out of our remit, what is the action that we can take as researchers to move closer to the foundations of a new system, and how would that shift our stakeholders' practice? If we know that the current foundation is unsuitable to be built upon, how do we get to a place where we can imagine a meaningful alternative in order to start building a new foundation?

Our work contributes to an extensive body of LIMITS literature that considers possible futures such as: what happens in futures where digital technology is constrained~\cite{tomlinson2017information, pargman2017resource, sutherland2021design}; designing for more inclusive futures~\cite{ahmed2016computing, thomas2018disableism}; future collapse scenarios~\cite{remy2015limits, penzenstadler2015collapse,jang2017unplanned, mcdonald20163d}; the impact of limits on systems complexity~\cite{raghavan2016refactoring}; and,  opportunities for policy and social change~\cite{joshi2016whose, houston2022richness}. Through our speculative design and reflections we  contribute to growing practices of activism and practice change at LIMITS. For example, critiquing normative practices like human-centred or user-centred design~\cite{thomas2017limits, dos2021we}, and reflections on activism in research practice~\cite{bates2020let}.

Speculative design is an approach to design that facilitates ``\textit{critical reflection through future narratives\ldots often mediated through objects}''~\cite{forlano2013ethnographies,dunne2013speculative}. In contrast to the more commercial forms of design that lead to the creation of the tools that our stakeholders use day-to-day, speculative design deals with the ambiguous and liminal spaces of the potentially fictional, seeking an understanding of ``\textit{hypothetical possibilities\ldots utopian concepts and dystopian counter-products}''~\cite{auger2013speculative}. Within the realm of speculative design, design fiction acts as a specific method
for building fictional worlds and ``\textit{exploring alternative visions of the future}''~\cite{kirman2022thinking}, often taking the form of the manifestation of alternative or future worlds ~\cite{blythe2014research,pargman2019future}, provoking conversations about e-waste~\cite{thomas2015more}, designing for less desirable and uncomfortable climate futures~\cite{kuijer2024feeling}, the exploration dystopian futures and collapse~\cite{tanenbaum2016limits}, participatory design fictions for survivability~\cite{burnell2018design}, and, counterfactual histories to imagine computing futures~\cite{eriksson2018meeting}.




\section{ANCSTRL.LAB, a human-centred energy futures consultancy}

\begin{figure}[h]
    \centering
    \includegraphics[width=8cm]{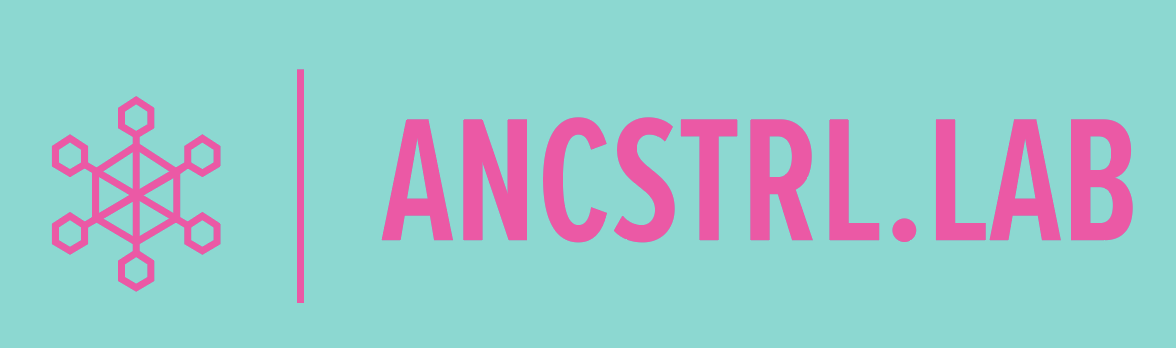}
    \caption{The logo for ANCSTRL.LAB---an human-centred energy futures consultancy from the future.}
    \label{fig:ancstrl-logo}
\end{figure}
We present our speculative designs through the window of ANCSTRL.LAB, a design fiction consultancy from the near future that plays with the idea of design consultancies as self-proclaimed agents of organisational transformation and architects of the future. They employ systems thinking, games, play, and other more-than-human approaches to help organisations rethink their energy saving decisions around the heating and cooling of buildings. These speculations aim to world build around the integration of systems change (cf.~\cite{meadows1999leverage}) and more-than human centred-design (cf.~\cite{coulton2019more}) into energy management practice.
ANCSTRL.LAB exists in an imagined reality where things are different. This gives us as designers freedom to push boundaries and consider systems change as reality in our designs. Design fiction allows for exaggeration and caricature~\cite{blythe2014research}.

What would tools that foreground more holistic thinking immersed in systems complexity look like from a radical, responsible, human-centred consultancy that actively wants to create systems change that may one day make their own practice and organisation obsolete? It is here that we situate our three speculative designs or `provocations' grounded in themes and emergent discussions from our research in commercial energy management. By doing so we want to imagine how people are doing the work in the future, focusing on what engaging with the ways in which people are part of complex systems, and how this could shift the mental models and imaginaries of energy management.

\subsection{Humane Energy Management Handbook}

\begin{figure}[h]
    \centering
    \includegraphics[width=8cm]{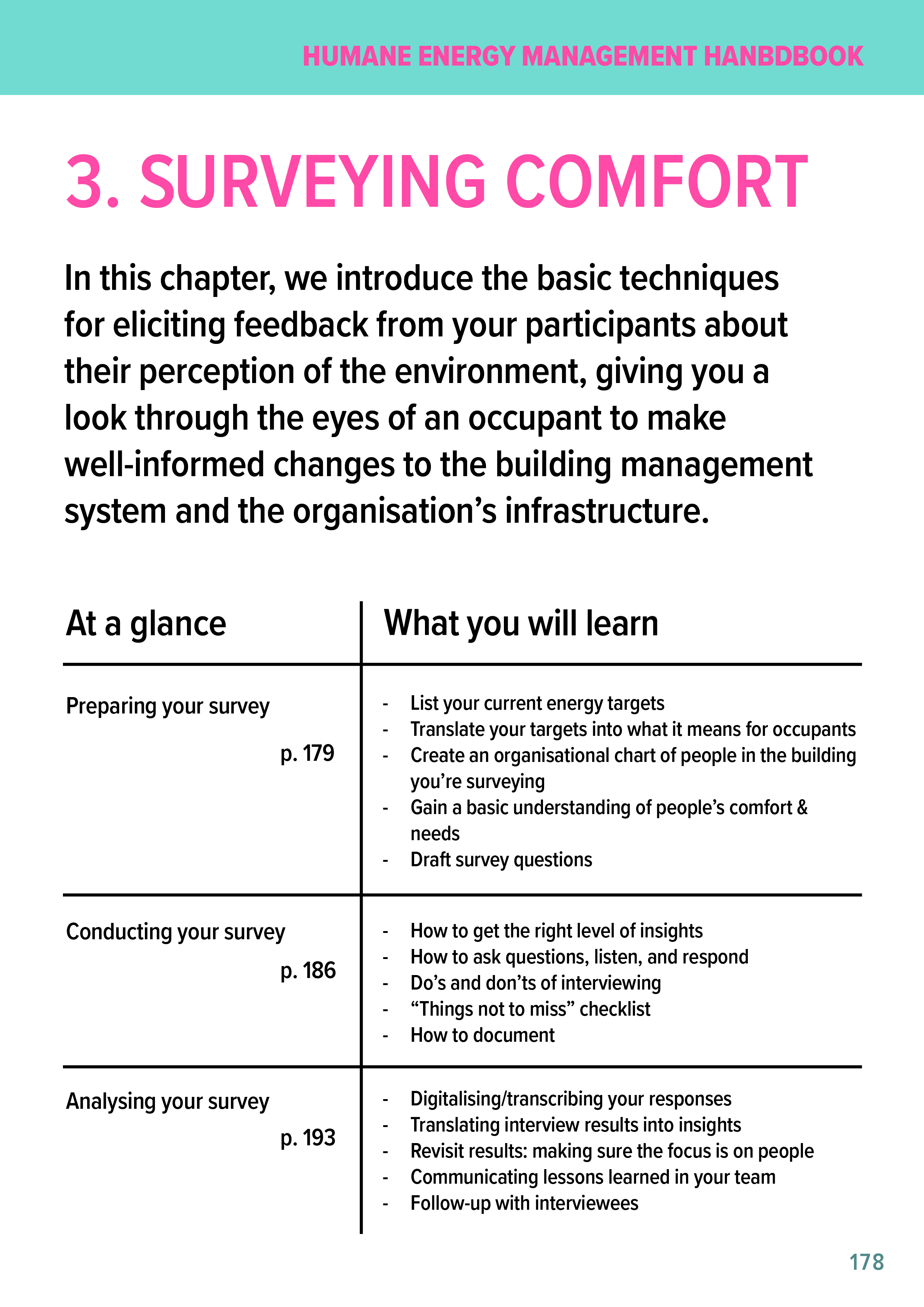}
    \caption{Humane Energy Management Handbook: The energy manager works through a 260 page handbook where they find a series of examples and exercises to help prepare and conduct surveys. In chapter 3 the energy manager will consider occupant experiences and feedback that can be used in the management of buildings to better serve occupants whilst reduce energy consumption of heating and cooling systems.}
    \label{fig:humane-handbook}
\end{figure}

\emph{Humane Energy Management Handbook (HEMH) (Figure~\ref{fig:humane-handbook} is a self learning guide and CPD (Continuous Personal Development) for energy stakeholders looking to build a human-centred approach into their energy management practice. This book reduces the gap between energy management practices and human experiences, giving energy stakeholders structure for gathering human context and including it in energy analysis. ANCSTRL.LAB provides CPD certified Humane Energy Management Training for groups and individuals upon request.}

In our project we set out to investigate energy savings, amongst other goals, in complex commercial settings, based on extensive data analysis from building management systems and other data sources. Just looking at the data is not sufficient to understand why an energy system anomaly is occurring or the impacts of energy reduction interventions. Talking to building users, caretakers or office managers is inevitable to get the whole story, be it to explain anomalies or patterns found during the data analysis, confirming that the status quo is maintaining a good comfort level for occupants, or uncovering opportunities for change. But while data is always available at the click of a button---at least in theory, barring hiccups in the data pipeline that are all too common---people are short on time. During the course of our project we encountered this as one of the obstacles to investigating energy savings: scheduling meetings with anyone involved in the complex hierarchy of the organisational setting pertaining to energy management. Even if a meeting went ahead, there were often questions that could not be answered, pieces of information that could not be provided, follow-up investigations that could not be launched because of lack of time. In contrast, quantitative data analysis tasks seem not to fall into this line of argumentation, even though they could end up yielding far less interesting insights, if any.

\begin{figure*}
    \centering
    \includegraphics[width=\textwidth]{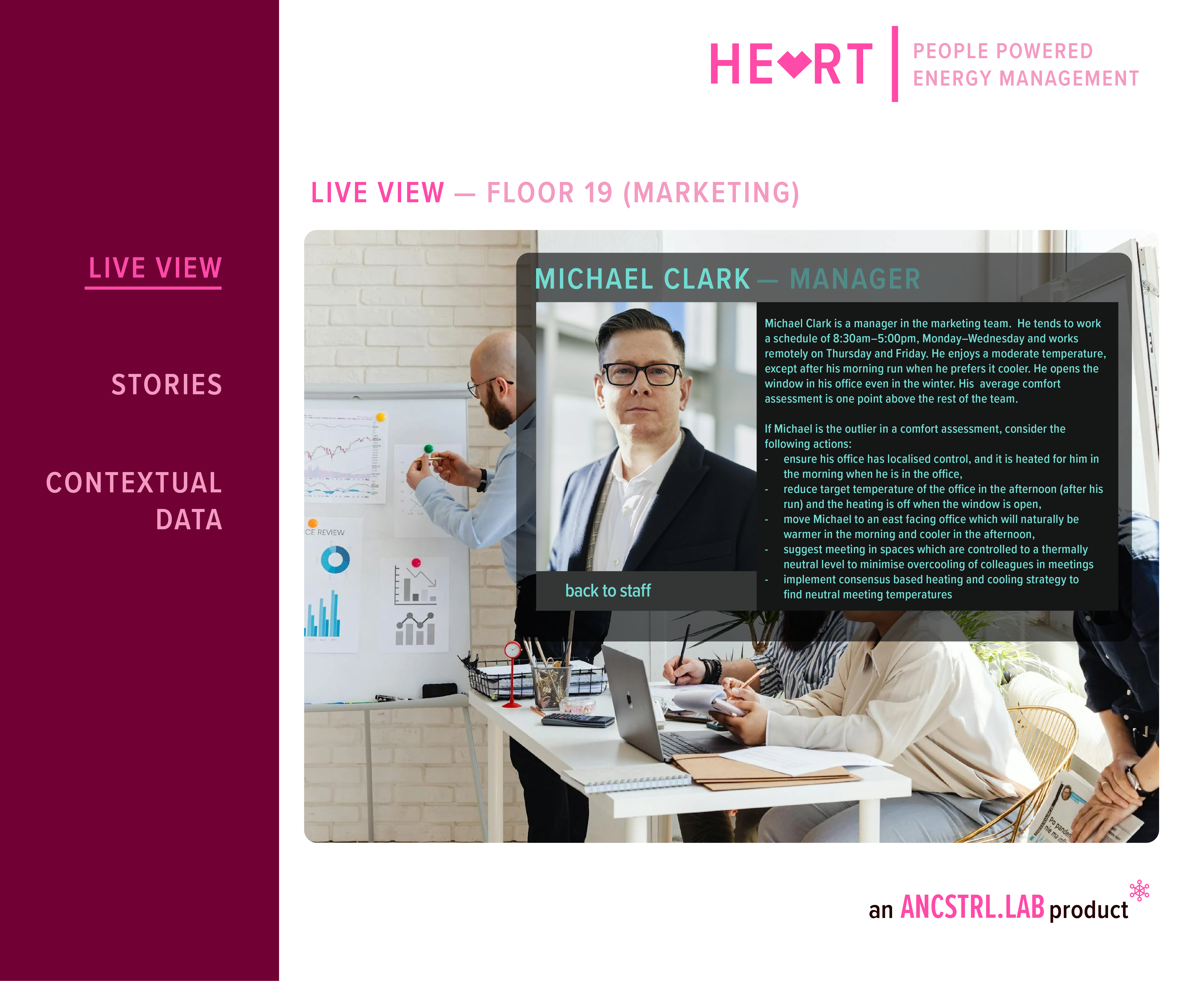}
    \caption{In the live view, the energy manager identifies the current occupants of a room and contextualise their typical energy usage behaviours. The energy manager clicks on an individual person, they are shown an explanation of the occupants typical energy usage patterns and any contextual factors. The live view explains that Michael Clark enjoys a cooler temperature after his morning run. This provides deeper context on the person behind the energy behaviour, and can help energy managers to act with a more granular understanding of the system they’re working within.}
    \label{fig:pp-em}
\end{figure*}

This highlights a core problem of the perception of time it takes for qualitative compared to quantitative data gathering and analysis, but it is not necessarily surprising. Energy management career paths cover a range of areas relating to energy systems. From degree level training, that focuses on methods and tools for measuring, analysing and optimising, one can become an expert in building management, energy demand, district heating, and a range of other complex sub-systems (e.g., lighting, solar generation) and areas relating to energy systems. These pathways will likely lead from analyst and technician roles to a senior management or consultant role that requires more holistic (or systems) thinking, including organisational structures, the needs of employees and building users, and the development of policies. A myriad of other possible pathways to energy management exist that may take into account human lived-experiences and needs, but this has not been our observation of the dozens of energy management stakeholders we have worked or engaged with. These pathways build particular skills and competencies, covering a narrow subset of areas which focus on measurement and optimisation and less on the elements relating to people such as human experiences, and organisational needs. We think a more generalist perspective is required to consider people, policy and the organisation.





This is all compounded by training and handbooks, such as the Energy Information Handbook~\cite{granderson2011energy}, that focus on quantitative data analysis almost exclusively. This form of training and development for energy managers largely entails dashboards, charts, rankings, and benchmarks. And the most common tool in the hands of energy managers we have worked with were by far and large spreadsheets. As one energy manager told us repeatedly in an interview about what to do with all the data: ``\textit{Context is king}''; but alas there was no strategy or process on how to gather context, or how it would be included in any energy analysis. People use buildings in diverse and sometimes complex ways. A more holistic and systems-thinking led approach to data contextualisation is needed to realise energy interventions with increase leverage. The HEMH does this through providing energy stakeholders with tools for contextual information gathering and analysis.   


\begin{figure*}[h]
    \centering
    \includegraphics[width=\textwidth]{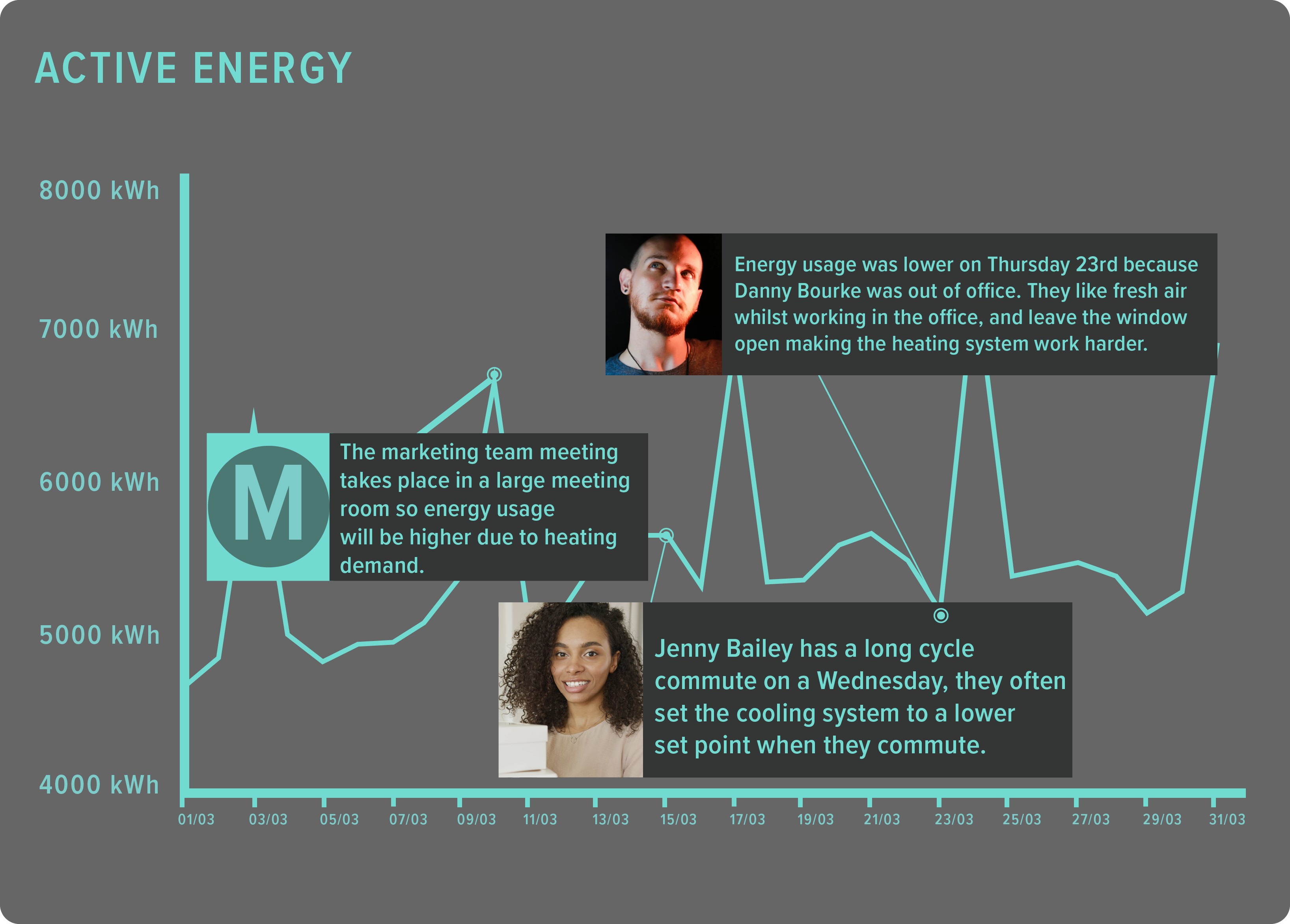}
    \caption{The energy manager opens the contextual data tab. As they look across the energy consumption for the the past month anomalies and patterns are annotated with occupant and business context. The energy manager views this extra context about the marketing team, Jenny and Danny, quickly sense checking the energy consumption for those days. With this context they have more information about the energy variation, and are content that the variation they see before them doesn't require further investigation.}
    \label{fig:cropped}
\end{figure*}








\subsection{Putting occupant context into Energy Management}

\emph{Human occupants use technology and spaces in ways that energy stakeholders might consider abnormal. We want to challenge the default lens of energy management which focuses on accounting and key-performance indicators (KPIs). This new dashboard tool (Figure~\ref{fig:pp-em}) presents an interface for energy stakeholders that is human-centred view by default, rather than leading with data that describes energy system performance. This can result in a chain of enquiry that is agnostic of human needs and experiences; that follows the numbers, focuses on optimisations, fact finding, and gathering growing quantitative time-series and quantitative data.  In this interface, the default view is that of building occupants, their activities, and warnings that challenge default data-driven approaches to intervention. The users can view time series energy analytics through views (Figure~\ref{fig:cropped}) that contextualise possible anomalies with human-centred information that wouldn't be surfaced in other traditional interfaces.}

\begin{figure*}[h]
    \centering
    \includegraphics[width=16cm]{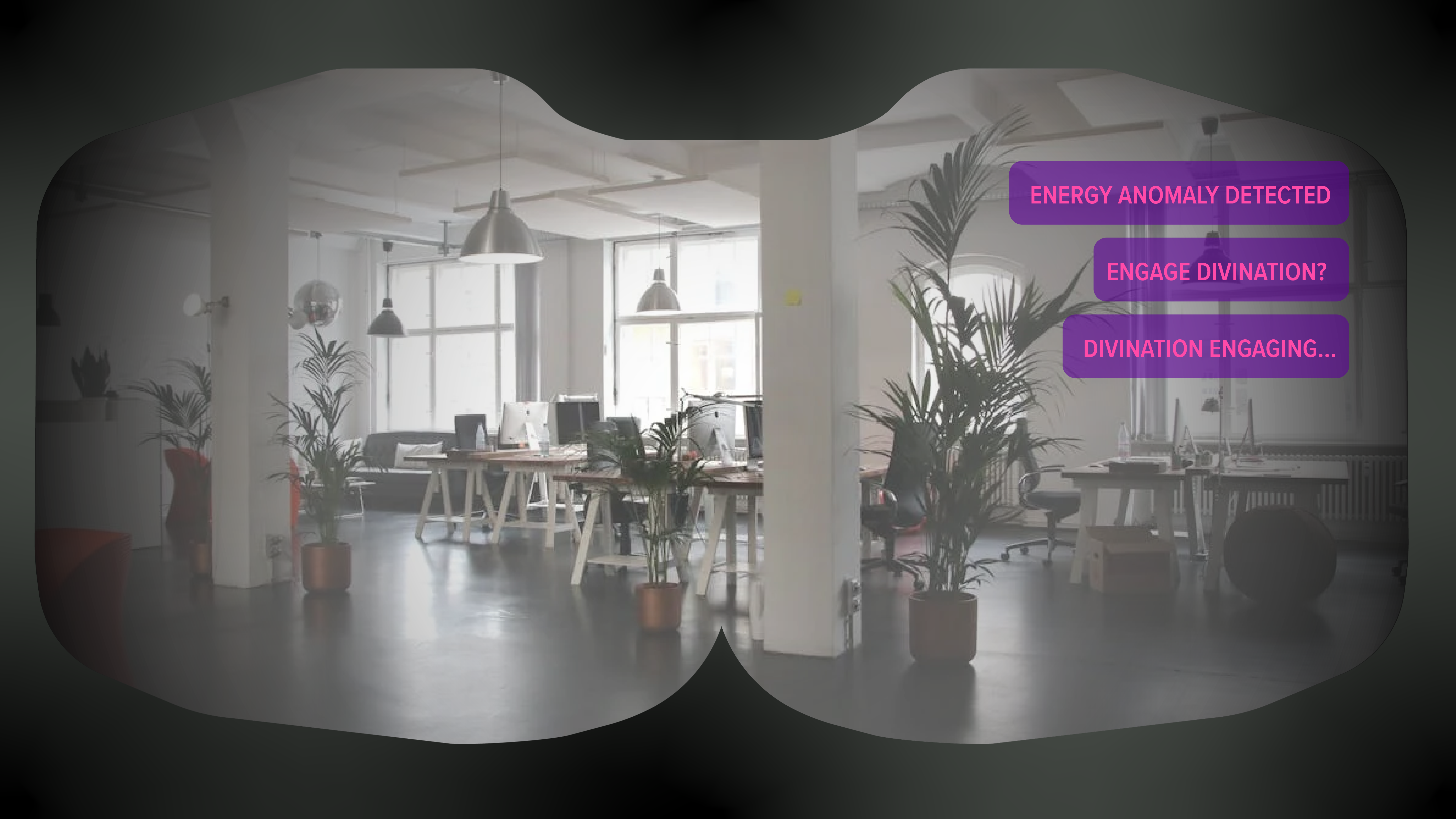}
    \caption{The energy manager puts on ContextSense, augmenting their vision with a spatial computing overlay. ContextSense immediately detects an anomaly, displaying the message, ``ENERGY ANOMALY DETECTED''. The wearer is prompted with the message ``ENGAGE DIVINATION''. Under their breath they utter the words ``yes'' as they brace for the intensity of flowing through the wake of blended supernatural energy and building data. The user is shown the message ``DIVINATION ENGAGING'', and ContextSense seamlessly attaches to all data feeds and building memories, and begins to process.}
    \label{fig:contextsense1}
\end{figure*}

\begin{figure*}[h]
    \centering
    \includegraphics[width=16cm]{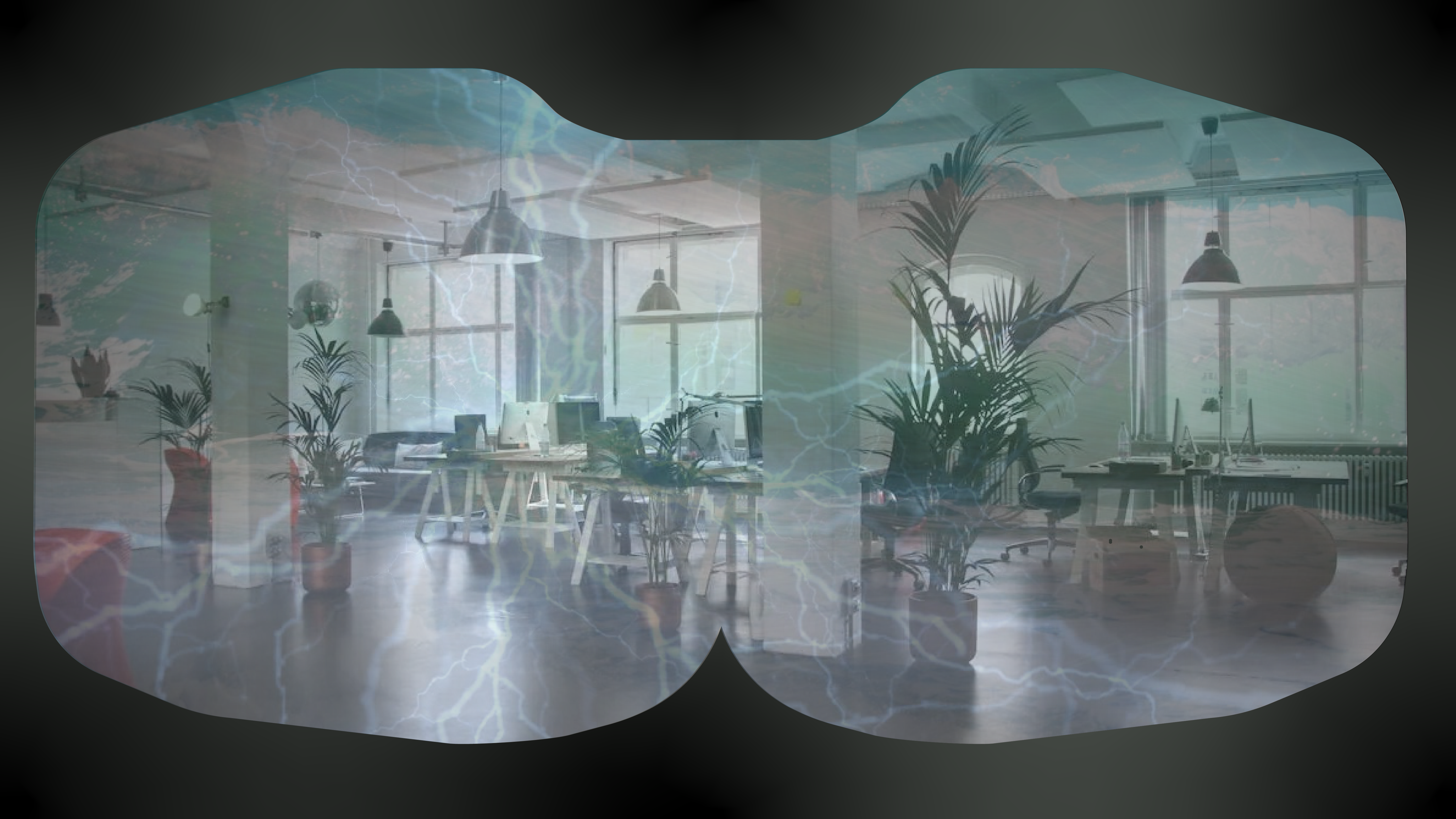}
    \caption{ContextSense draws their focus to the desks through a display of energy intensity that looks like sparks and lightning. The wearer awaits a following visual clue to the source of energy intensity.}
    \label{fig:contextsense2}
\end{figure*}

Dashboards and spreadsheets seem to be the default starting point for energy management stakeholders. A wide range of energy management data dashboards can be procured to help commercial organisations make more data-informed decisions about their energy systems. Throughout our work with energy managers, energy consultants and those responsible for energy policy, we are often met with visualisations, dashboards, and excel spreadsheets. Our partners were primarily using Microsoft Excel to analyse energy and building data. Whilst Excel does have relatively advanced features for mathematical functions, it is an accounting tool. It is not designed for granular time-series data or to capture and communicate rich qualitative data alongside KPIs and other quantitative analyses. The scale of such data grows proportional to the size of a commercial estate, and the assets (e.g., controllable devices, measurable things) that are contained within said estate. Secondary data such as weather data is also required for energy management analyses such as degree day analysis, a technique for normalising energy consumption in relation with outdoor temperature. From our engagements with energy stakeholders we have developed data quality and anomaly detection modules for dashboards to help stakeholders reduce the noise in data, fix poor quality data feeds and take action on perceived anomalies in energy consumption. We have come to understand that the noise is a function of poor data collection or storage, whilst anomalies that are not related to data quality are due to normal human actions in service of the commercial or business interests of the organisation, such as using space heaters to keep warm in cold offices with poorly running heating infrastructure, leaving chiller doors open whilst loading fridges, or running energy intensive ICT hardware heating spaces in unexpected ways. Without considering building use and business activities in energy management interventions there is a risk of negative impacts on users, productivity, and profit.

This speculation provides a mapping of people to energy infrastructure and possible energy interventions. These interfaces can help push beyond the default mindsets of energy management that focus on maintenance of infrastructure or reduction of energy. Through more person-centric reporting and suggestions and the mapping of building user and organisational context on top of energy data it's possible to consider interventions that consider more aspects of the organisational systems, such as the flexibility of people and spaces, and opportunities for radical new energy policy.

\subsection{ContextSense}

\emph{It's difficult to make decisions when considering the needs of people, buildings and organisations: having a technology that gives you an intuitive sense of clarity helps you see the whole system and all its complexity. ContextSense (Figures~\ref{fig:contextsense1} and~\ref{fig:contextsense2}) is a unique spatial computing headset for energy management driven by AI augmented reality (AR) technology (patent pending). It provides the user with the power to slip into augmented realities where our AI influences and augments the users vision with lucid visual representations divined from all signals contained within or connected to a building. These representations require the user to interpret and project, giving the user the power to gather holistic signals drawn from sick energy systems, ethereal occupants and the memories of buildings. By harnessing both building data and supernatural energy our AI speaks in visual riddles that must be navigated and mastered. Upon mastery the user can foreground context and navigate systems complexity, whilst extracting future consent for interventions whenever needed. This technology helps users regale convincing stories about the past, present and future of a building, its users and the energy entangled within.}


Energy management has a range of levers to affect changes in an organisation. We observed
a discrepancy between actions taken and the potential impact that can be achieved through an intervention. The most striking examples are signs in utility rooms and liminal spaces, such as bathrooms, kitchens, meeting rooms, or hallways, to switch off lights; at the same time, the very same spaces are equipped with energy-consuming `always on' appliances such as energy intensive public displays, air flow vents, or IoT devices---some of which consume orders of magnitude more than a ceiling light left on, especially considering the energy efficiency of LED lights. This also highlights the discrepancy (and hypocrisy) between blaming building users (`switch off the lights') versus intervening in more systemic ways. A more systemic intervention in this case could consider automatically switching off appliances in empty room, revisiting whether information needs to be displayed on a public screen or an analogue sign, and adjusting ventilation settings to levels appropriate to occupancy levels or occupant preferences. These kinds of interventions require both a mindset shift in energy management practice and better signals to energy management about more qualitative or emotive aspects of building use such as the emptiness of spaces, changing occupancy, comfort of individuals and groups, and the effectiveness of information displayed in public spaces.

For energy managers there is another problem: the most impactful change to increase a building’s energy efficiency is to make substantial changes to it (insulate, renovate, retrofit, or worst case even rebuild it). This can make stakeholders feel like their interventions are just fixes with limited impact, likely to be temporary, and unable to intervene at the root cause of an issue. At the same time, the changes they can make---such as smaller infrastructural investments, or changing settings in energy system controls---commonly have different levels of costs and impacts. Whilst changing light bulbs to more energy efficient LEDs is a relatively quick and low-cost fix, its impact is rather small, and pales in comparison to the energy spent on heating. Intervention such as these have to be planned whilst in conversation with occupants and/or organisational management, which might include multiple rounds of discussions, gathering feedback, and accruing consent for investment.

ContextSense perhaps is a double-edge sword. It can give the user ultimate power over future energy interventions due to the way it navigates complexity and decision making processes regarding  buildings, energy, and organisational structures. If such a technological intervention considers all aspects of a complex system and can divine an appropriate changes, what does that mean when balance is found? Can such a technology change the rules of a system or the whole paradigm by reaching into possible futures? Or will it lead to chaos and the corruption of the user?

\section{Discussion}

Through the creation of our design fiction and speculations we have developed new ideas about how to communicate and visualise complex systems, especially when working in data-heavy, engineering oriented spaces. In this section we discuss our reflections and the opportunities for speculation and design fiction practice at LIMITS.

\subsection{What have we learned?}
\subsubsection{Playful speculations}

Energy management practice is often serious, entirely quantitative, and utilitarian. We were interested in bringing more creative and playful methods and ideas into this space. ContextSense, for example, builds on the idea of artefacts and playful ways to prompt and probe for context in situ. The tabletop game Mysterium\footnote{Mysterium is a co-operative deductive game where a group of players work together to decipher supernatural `visions' sent from a ghost to solve murders.} and tarot are influences where visual tools (cards) are used to bring out reflection with few or no words, and certainly no quantitative data. This is a provocation that speaks to the data heavy nature of energy management, as well as the intuition that experienced energy stakeholders use when attempting to fix energy systems and work with building users and organisational management.

Design fiction and speculative design are creative ways of imagining desired futures and the changes we wish to drive towards. We believe that these approaches provide LIMITS researchers with a tool that can help with world-building, the construction of new imaginaries (or mindsets) with a range of stakeholders, and help us hold spaces where systems change can be made possible. Like Easterbrook, we see playfulness, games, and systems thinking as highly compatible for bringing people on a journey and telling them stories about possible changes to systems~\cite{easterbrook2014computational}, which can help change mindsets and help stakeholders to think about systems.

\subsubsection{Using new tools and shifting mindsets}


Building on the calls for more sociological perspectives in commercial energy management (cf.~\cite{gregg2024getting}) we see speculative design as a useful approach for participatory world building (cf.~\cite{burnell2018design}) with a broad range of stakeholders and perspectives, perhaps even reaching the mindsets of policy makers and executives. By working between sociology and speculative design, we can explore the ``\textit{common byways}'' between the disciplines in ``\textit{open, multiple, uncertain and playful ways}''~\cite{michael2012designing}. This kind of speculative work helps to better contextualise energy management practice in the surrounding complex systems, how existing tools used and taught to professionals do not account for humans or their experiences, and that the existing system is to blame for energy management practice's current shortcomings.
When designing interventions around energy data and carbon reduction, we have found ourselves repeatedly draw to designing for the lenses and tools that energy managers and energy stakeholders primarily use. Whilst using these lenses and tools, we are restricting ourselves to design within the existing mindsets of the stakeholders. Using design fiction has helped us shift our mindsets about the possible future tools and approaches that could change energy management practice.


Energy management practice is constructed entirely within the paradigm of neoliberal capitalism. Energy management practice is intertwined with ``\textit{engineering-led tools [that] are being mobilized to pursue profits in the name of carbon reduction}''~\cite{gregg2024getting}, continuing to play out Jevons paradox of efficiency improvements leading to increased use (cf., ~\cite{alcott2005jevons}). It can seem impossible to imagine an alternative to the current forms of energy management practice that are centred on capitalism~\cite{fisher2009capitalist}. The practice and tools of energy management is where our research is situated, and as such is where we can encourage more systems thinking (e.g., through more human-oriented approaches~\cite{gregg2024getting, alhamadi2022data}), which themselves limit the consideration of non-humans and ecology (cf.~\cite{thomas2017limits, dos2021we})

In our prior work, and the work of the Net0i project more broadly, we focus on suggestions that target increased efficiency and cost saving which are practical in the current configuration of the capitalist system, calling for policy interventions, and for technology developers to consider the materiality and energy of their products or services. Yet in this paper, we have explored design fiction as a way to more deeply explore the alternatives that are difficult to imagine. We see design fiction as a flavour of LIMITS research that can help us all imagine more clearly what might be possible if things start to change.  In turn, this helps new ideas, movements, focus, to be in-becoming, allowing ourselves permission to shift our mindsets, and just start the creation of a thing that is possible.




\subsection{Moving from critique of the consultancy model to speculative praxis for systems change}

Through design fiction we have wanted to surface possible ideas that could drive systems change in energy management practice. Through the development of the design fiction and speculations we have been drawn to the following reflections on possible ways (for us, and possibly you) to move from a place of critique to a praxis, a politics (informed by theory and critique) that can be actioned~\cite{deleuze2004desert}.
ANCSTRL.LABS was designed to provoke friction with the ideas of current management and technology consultancies who are procured to help leverage particular outcomes or changes to an organisation. This has been described as ``\textit{buying outcomes}'' where organisations are only paid upon successful creation of outcomes ~\cite{cfpi2024}. This sits in tension with organisational systems, as changes in these outcomes only arise as emergent properties of of a complex system rather than as a transaction. The views in this discussion are situated in our experiences as researchers and consultants who have worked with medium and large commercial organisations who predominantly focus on view of energy and sustainability through the lens of  KPIs and metrics. We would like to call out that not every design or energy consultancy operates in a predominantly neolibreal context where energy is considered through a commercial frame, and that there are a range of mission driven organisations (consultancies) working with communities in areas such as energy security and poverty, cooperative ownership, and localised energy solutions that do not sit in this framing. 


The current model looks like this: the client (medium and large organisations) procures the services of consultancy firms to consult on high-level, `impossible' and complex `problems'. The responses to these problems manifest as products (or solutions?) that are top down product-, service-, or organisational strategy-level transformations. Practically this comes in the form of a report (and presentation) that couples a range of solutions (that the consultancy can often provide) to the clients' key performance indicators (KPIs). In the context of commercial energy management this could look like: 1) encouraging an organisation to use technology and agile processes to make it less resource intensive through the reduction of waste energy, and, 2) re-imagining client's energy management strategy or policy to find interventions that help them utilise emerging market trends (e.g., AI, digital twins) or respond to policy/regulatory changes (e.g., Net Zero).

The outputs from the consultancy (evaluation reports, spreadsheets, slide decks) are used to justify and initiate changes within an organisation, in order to attempt to create changes in outcomes. If a higher up in an organisation wants to change the way an organisation functions, they may procure a consultancy, who gather primary and secondary data (through business analysis and user centred design), who prototype solutions grounded in this data, and then demonstrate the potential impact of their solutions on the client organisation. These kinds of solutions often implicate the consultancy in future paid work.

Consultancies act and are treated as experts and as such, key strategic business decisions are outsourced to them. This is a form of intellectual and strategic capital accumulation, limiting their clients from autonomously thinking about or doing things differently the future. The lack of diverse future possibilities for energy management is in lockstep with what Fisher describes as Capitalist Realism, ``\textit{the widespread sense that [\ldots] it is now impossible even to imagine a coherent alternative to [capitalism]}''~\cite{fisher2009capitalist}. For our experiences, in the minds of commercial energy management the possible future is one that focuses on using data to reduce financial and carbon costs, meeting regulatory requirements today, and ensuring that growth (of business, of profits) can continue in the present. These mindsets are tinged with some hope that technology will solve today's problems sometime in the future (cf.\ the lack of possible alternatives to capitalism in Capitalist Realism~\cite{fisher2009capitalist}). This is where the consultancies come in. They promise innovative solutions that their clients cannot comprehend or achieve alone. This reliance on the consultancy as the solution provider gives consultancies immense power over clients, keeping them tightly aligned, dependent and locked-in to the limited range of possible futures that consultancies create---futures that these consultancies may not even know how to move closer towards. Further reinforcing the flow of capitalist realism is the consultancy who are never going to suggest degrowth or models of post-capitalism/post-neoliberalism or co-operation because these concepts are in direct conflict with their own future survival and growth.

Part of the currency in this system (between client and consultancy) is often evaluation reports containing a series of recommendations for the client organisation. In the context of commercial energy these consultancies are being paid to tell us how to manage our energy. They offer a report and their expertise as a token that helps organisations make incremental changes and reduce expenditure. All while sustaining the existing system and limiting the range of possible futures as clients become increasingly dependant on consultations like this and locked-in to an increasingly narrow range of future possibilities. This means that the evaluation report (or technology interventions) are the token by which change to a systems is or is not made.


\emph{What if energy managers actively connected energy consumption and human experiences in ways that focus on comfort and joy, rather than reducing energy bills? Other than dashboards and spreadsheets, what default interfaces could energy management practitioners use? How resilient are energy management mindsets when responding to rapid emergent changes to a system? Is energy management prepared for climate devastation?}

\subsubsection{From speculative design to Speculative Praxis}


The act of diegesis, is the imparting or communicating of information through fictional works, in our case a design fiction and speculations. By drawing on speculative design, design fiction, and diegetic prototypes, we can consider technologies for ``\textit{implausible social goals}''~\cite{gaver2000alternatives} focusing on practices that prefigure the worlds we are trying to build~\cite{asad2018prefigurative}, rather than limiting ourselves to plausible and practical technological solutions~\cite{coulton2017design}. As we have demonstrated, speculative design can be used to help elicit and communicate ``\textit{experiences, frustrations, and joys}'' of a system~\cite{kirman2022thinking}, making the fiction more compelling and playful.

The creation of diegetic prototypes develop the fictional positioning suggested by a design fiction and enhance their realism by ``\textit{creating a full elaboration of the technological diegesis which includes any part of the fictional world concerning the technology}'' ~\cite[p. 46]{kirby2010future}. As such, diegetic prototypes work to effectively represent pieces of imagined, fictional worlds and make them seem and feel more possible to their audiences through a more direct engagement with the world. In so doing, they also create new possibilities and capacities for those engaging with them. An example of this is presentation of design fiction in a public policy brief helping policy makers imagine possible policies that considers e-waste in procurement, a policy that was unfathomable until it was conceptualised as possible in the design fiction ~\cite{thomas2017hci}\footnote{See the Supplemental Material, Appendix B for Design Fiction Press Release}.

When realist and normative approaches to technology design fail to impact a system because of their lack of ability to challenge mindsets and paradigms, these methods can create new imaginaries, which in turn can enable the questioning of existing mindsets and paradigms (cf.~\cite{meadows1999leverage}). The bringing of diagetic prototypes to stakeholders in a design or research activity functions as an activist ``\textit{speculative praxis}''~\cite[p. 128]{cutting2022towards}, through which speculative design activity can begin building a different material reality. We see speculative praxis as a necessary approach for encouraging post-neoliberal interventions, technologies, and organisations in commercial energy and beyond. Whilst the fall of neoliberal capitalism and the possibility of any alternative may seem unfathomable, it is necessary that we find alternatives to help reduce ecological devastation and climate change that will have unknowable impacts on the lives of humans, nature, and all living beings. Through the new mindsets and paradigms that speculative praxis can make fathomable. we can start to encourage our ideas about possible different futures fractal out from the small scale (e.g., our own research practice), to how we wish for things to be at a larger scale.

\begin{quote}
    ``\textit{How we are at the small scale is how we are at the large scale. The patterns of the universe repeat at scale. There is a structural echo that suggests two things: one, that there are shapes and patterns fundamental to the universe, and two, that what we practice at a small scale can reverberate at the largest scale.}''~\cite{maree2017emergent}
\end{quote}






\section{Conclusion}
In this paper we have developed a series speculative designs to reflect on the Net0i project and explore where systems change and design fiction come together.
We have presented ANCSTRL.LAB as a fictional consultancy with a range of products and services from the future. Heart, the Humane Energy Management Handbook, and ContextSense all strive futures where human contexts are integral in commercial energy management practices. Our discussion considers the use of play and the shifting of mindsets when designing in LIMITS contexts.
The regular paradigm for energy managers focuses on quantitative data,
KPIs and under-baked ideas of systems. Through design fiction we drive at the need for deeper acknowledgement by energy stakeholders of complex systems that include people. We have finished by articulating how speculative praxis could take these ideas one step further, as a potential way of changing mindsets through the building of new material realities.


\begin{acks}
We thank our funder, EPSRC (Grant ref.\ EP/T025964/1), Oliver and Kieran thank their colleagues and friends Leah and Hazel from fractals co-op for the support and encouragement. 
\end{acks}

\bibliographystyle{ACM-Reference-Format}
\bibliography{thebib}
\balance








\end{document}